 \documentclass[final,3p,times,sort&compress]{elsarticle}

\usepackage{multirow,setspace,amssymb,amsmath,graphicx,color,rotating,subfigure,url}
\usepackage{lineno}
\usepackage{natbib}
\usepackage{textcomp}
\journal{Physica A} 

\begin{document}

\begin{frontmatter}

\title{Statistical properties of visibility graph of energy dissipation rates in three-dimensional fully developed turbulence}

\author[BS,RCSE,RCE]{Chuang Liu}
\author[BS,RCSE,RCE,SS,CAS]{Wei-Xing  Zhou\corref{cor}}
\cortext[cor]{Corresponding author. Address: 130 Meilong Road, P.O.
Box 114, School of Business, East China University of Science and
Technology, Shanghai 200237, China, Phone: +86 21 64253634, Fax: +86
21 64253152.}
\ead{wxzhou@ecust.edu.cn} %
\author[RCSE,SKLCE]{Wei-Kang Yuan}

\address[BS]{School of Business, East China University of Science and Technology, Shanghai 200237, China}
\address[RCSE]{Engineering Research Center of Process Systems Engineering (Ministry of Education), East China University of Science and Technology, Shanghai 200237, China}
\address[RCE]{Research Center for Econophysics, East China University of Science and Technology, Shanghai 200237, China}
\address[SS]{School of Science, East China University of Science and Technology, Shanghai 200237, China}
\address[CAS]{Research Center on Fictitious Economics \& Data Science, Chinese Academy of Sciences, Beijing 100080, China} %
\address[SKLCE]{State Key Laboratory of Chemical Engineering, East China University of Science and Technology, Shanghai 200237, China}

\begin{abstract}

We study the statistical properties of complex networks constructed from time series of energy dissipation rates in three-dimensional fully developed turbulence using the visibility algorithm. The degree distribution is found to have a power-law tail with the tail exponent $\alpha=3.0$. The exponential relation between the number of the boxes $N_B$ and the box size $l_B$ based on the edge-covering box-counting method illustrates that the network is not self-similar, which is also confirmed by the hub-hub attraction according to the visibility algorithm. In addition, it is found that the skeleton of the visibility network exhibits excellent allometric scaling with the scaling exponent $\eta=1.163\pm0.005$.

\end{abstract}


\end{frontmatter}

\section{Introduction}
\label{S1:Introduction}

Turbulence is one of the most challenging fields in physics, and
diverse methods have been adopted to investigate the statistical
properties of turbulent flows
\cite{Frisch-1996,Bohr-Jensen-Paladin-Vulpiani-1998}. For instance,
the multifractal nature of turbulent flows has been extensively
studied based on the structure function analysis of velocity time
series \cite{Anselmet-Gagne-Hopfinger-Antonia-1984-JFM} and the
partition function analysis of energy dissipation rates
\cite{Mandelbrot-1974-JFM,Meneveau-Sreenivasan-1991-JFM}. Due to the
extreme complexity of turbulence, phenomenological investigation of
experimental data plays a crucial role in order to gain a better
understanding of turbulence, which is fundamental for the
construction of models and theories. In this work, we provide a
first attempt to study turbulent signals from the complex network
perspective, hoping that the ideas, tools, and theories from the
flourishing field of complex networks can stimulate the study of
turbulence from an alternative point of view. It has
been shown that the idea of complex network analysis is able to
identify flow patterns and nonlinear dynamics of gas-liquid
two-phase flows \cite{Gao-Jin-2009-PRE,Gao-Jin-2009-Chaos}.

Recently, several methods that convert time series into networks
have emerged. Zhang et al. introduced a method to deal with the
pseudoperiodic time series and found that the structure of the
corresponding network depended on the dynamics of the series
\cite{Zhang-Small-2006-PRL,Zhang-Sun-Luo-Zhang-Nakamura-Small-2008-PD}.
In this method, the nodes correspond directly to cycles in the time
series, and the edges are determined by the strength of temporal
correlation between cycles. This method for pseudoperiodic time
series can also be generalized to other time series, where a node is
defined by a sub-series of fixed length as an alternative of a
cycle, which has been applied to stock prices
\cite{Yang-Yang-2008-PA}. Xu et al. proposed another method, which
embeds the time series into an appropriate phase space, takes each
phase space point as a node in the network, and connects each point
with its four nearest neighbors to form a complex network
\cite{Xu-Zhang-Small-2008-PNAS}. There are also other mapping
methods from time series to networks based on n-tuples of
fluctuations
\cite{Li-Wang-2006-CSB,Li-Wang-2007-PA,Kostakos-2009-PA} or
recurrence plots
\cite{Marwan-Donges-Zou-Donner-Kurths-2009-PLA,Donner-Zou-Donges-Marwan-Kurths-2010-NJP}.

Lacasa et al. proposed the visibility graph algorithm, which can map
all types of time series into networks
\cite{Lacasa-Luque-Ballesteros-Luque-Nuno-2008-PNAS}. In the
visibility algorithm, the nodes correspond to the data points of the
time series in the same order, and an edge is assigned to connect
two nodes if they can ``see'' each other. It is found that the
degree distributions of visibility graphs converted from
self-similar time series have power-law tails
\cite{Lacasa-Luque-Ballesteros-Luque-Nuno-2008-PNAS} and the tail
exponents depend linearly on the Hurst index of the
original time series
\cite{Ni-Jiang-Zhou-2009-PLA,Lacasa-Luque-Luque-Nuno-2009-EPL}, in
which the multifractal nature of the time series has a negligible
influence \cite{Ni-Jiang-Zhou-2009-PLA}. The visibility algorithm
can be simplified to a variant termed the horizontal visibility
algorithm, which is solvable
\cite{Luque-Lacasa-Ballesteros-Luque-2009-PRE}. The visibility
algorithm has been diversely used to investigate stock market
indices \cite{Ni-Jiang-Zhou-2009-PLA,Qian-Jiang-Zhou-2009-XXX},
human strike intervals \cite{Lacasa-Luque-Luque-Nuno-2009-EPL}, the
occurrence of hurricanes in the United States
\cite{Elsner-Jagger-Fogarty-2009-GRL}, and foreign exchange rates
\cite{Yang-Wang-Yang-Mang-2009-PA}.

In this paper, we study the statistical properties of
complex networks constructed from the energy dissipation rate time
series using the visibility algorithm. In Section
\ref{S1:Data}, we describe the turbulence data set.
Section \ref{S1:VG} briefly describes the visibility graph
algorithm. The main results on the degree distribution, the
nonfractality and the allometric scaling of the
constructed visibility are given in Section
\ref{S1:Results}. We summarize in Section
\ref{S1:Conclusion}.

\section{Description of the data set}
\label{S1:Data}

The velocity data set from three-dimensional fully developed
turbulence has been collected at the S1 ONERA wind tunnel by the
Grenoble group from LEGI
\cite{Anselmet-Gagne-Hopfinger-Antonia-1984-JFM}. The mean velocity
of the flow is approximately $\langle{v}\rangle = 20 $m/s
(compressive effects are thus negligible). The root-mean-square
velocity fluctuations is $v_{\mathtt{rms}} = 1.7 $m/s, leading to a
turbulence intensity equal to $I = {v_{\mathtt{rms}}} /
{\langle{v}\rangle} = 0.0826$. This is sufficiently small to use
Taylor's frozen flow hypothesis. The integral scale is approximately
4m but is difficult to estimate precisely as the turbulent flow is
neither isotropic nor homogeneous at these large scales. The
Kolmogorov microscale $\eta$ is given by
\cite{Meneveau-Sreenivasan-1991-JFM}
\begin{equation}\label{Eq:eta}
    \eta = \left[\frac{\nu^2 \langle{v}\rangle^2}{15 \langle(\partial
   v/\partial t)^2\rangle }\right]^{1/4} = 0.195~{\rm{mm}},
\end{equation}
where $\nu = 1.5 \times 10^{-5} \rm{m^2 s^{-1}}$ is the kinematic
viscosity of air, and $\partial v/\partial t$ is evaluated by its
discrete approximation with a time step increment $\partial t =
3.5466 \times 10^{-5} \rm{s}$ corresponding to the spatial
resolution $\delta_{r} = 0.72 \rm{mm}$ divided by
$\langle{v}\rangle$, which is used to transform the data from time
to space by applying Taylor's frozen flow hypothesis. The Taylor
scale is given by \cite{Meneveau-Sreenivasan-1991-JFM}
\begin{equation}\label{Eq:lambda}
    \lambda =\frac{\langle{v}\rangle v_{\mathtt{rms}}}{\langle
   (\partial v/\partial t)^2 \rangle^{1/2}} =16.6~{\rm{mm}}.
\end{equation}
The Taylor scale is thus about $85$ times the Kolmogorov scale. The
Taylor-scale Reynolds number is

\begin{equation}\label{Eq:Re}
    {\rm{Re}}_\lambda = {\lambda v_{\rm{rms}}}/{\nu} = 2000.
\end{equation}
This number is actually not constant along the whole data set and
fluctuates by about $20\%$. We have checked that the standard
scaling laws reported in the literature are recovered with this time
series. In particular, we have verified the validity of the
power-law scaling $E(k) \sim k^{-\beta}$ with an exponent $\beta$
very close to ${5}/{3}$ over a range more than two orders of
magnitude, similar to Fig. 5.4 of \cite{Frisch-1996} provided by
Gagne and Marchand on a similar data set from the same experimental
group.

\begin{figure}[htb]
\centering
\includegraphics[width=7cm]{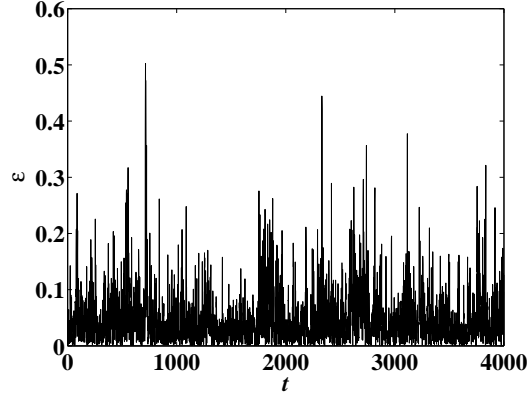}
\caption{\label{Fig:ts} A segment of 4000 data points of the energy
dissipation rate time series. }
\end{figure}

Then, we obtained the kinetic energy dissipation rate data using
Taylor's frozen flow hypothesis which replaces a spatial variation
of the fluid velocity by a temporal variation measured at a fixed
location, and the rate of kinetic energy dissipation at position $i$
is
\begin{equation}\label{Eq:eps}
    \epsilon_i \sim \left[ \left(v_{i+1} - v_i \right) / \delta_\ell \right] ^2
\end{equation}
where $\delta_\ell$ is the resolution (translated in spatial scale)
of the measurements. Fig.~\ref{Fig:ts} shows a randomly selected
segment of 4000 values of the energy dissipation rate time series.
The energy dissipation rate time series exhibits multifractal nature
\cite{Meneveau-Sreenivasan-1987-PRL, Zhou-Sornette-2002-PD}.

\section{Construction of visibility graph}
\label{S1:VG}

Here, we give a brief introduction of the visibility graph
algorithm. For a given time series ($t,\epsilon$), where $t$ is the
time variable and $\epsilon$ is the value of energy dissipation
rate, a visibility line exists between two points
($t_{a},\epsilon_{a}$) and ($t_{b},\epsilon_{b}$), if any other data
($t_{c},\epsilon_{c}$) placed between them fulfills:
\begin{equation}
\epsilon_c < \epsilon_b +(\epsilon_a-\epsilon_b)
 \frac{t_b-t_c}{t_b-t_a}~.
\label{Eq:VG_method}
\end{equation}

\begin{figure}[htb]
\centering
\includegraphics[width=7cm]{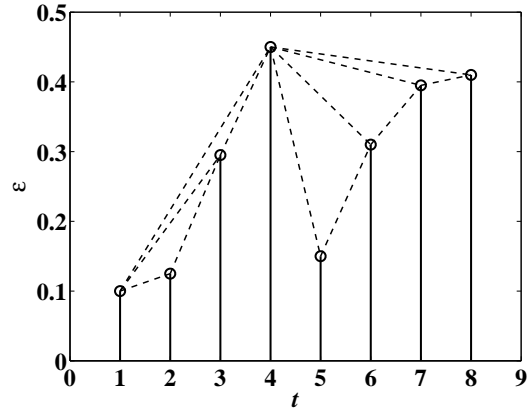}
\caption{\label{Fig:tsgraph} The visibility graph of a short time
series. }
\end{figure}

Fig.~\ref{Fig:tsgraph} illustrates the scheme of the visibility
algorithm, where eight points of the energy dissipation rate series
are plotted as an example. The height of the vertical lines are the
values of energy dissipation rates. The cycles and the dashed lines
constitute the visibility graph. The nodes correspond to series data
in the same order and an edge connects two nodes if one can be seen
from the top of the other. In other words, two nodes are connected
if there is visibility between them. Take points 1 and 4 of
Fig.~\ref{Fig:tsgraph} as the example to explain the concept of
``visibility''. Between 1 and 4, there are two points 2 and 3 which
are all under the dashed line from point 1 to point 4, and
visibility exists between nodes 1 and 4.

The networks mapped according to the visibility graph algorithm are
connected and undirected. Different time series convert into
different networks. It is easy to check that for the time series
with monotonously decreasing slopes of all the adjacent points, a
chain-like graph will be obtained.  According to this algorithm, the
dynamics of the time series is conserved in the graph topology. For
instance, Lacasa et al. found that periodic time series convert into
regular graphs, random time series into random graphs, and fractal
series into scale-free graphs
\cite{Lacasa-Luque-Ballesteros-Luque-Nuno-2008-PNAS}.

\section{Statistical properties of the visibility graph}
\label{S1:Results}

\subsection{Degree distribution}

\begin{figure}[htb]
\centering
\includegraphics[width=7cm]{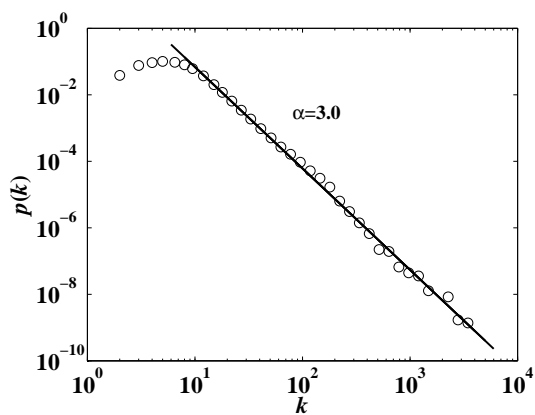}
\caption{\label{Fig:degreedistribution} The degree distribution of
the visibility graph converted from the time series of the energy
dissipation rate series. The line is the best power-law fit of the
data.}
\end{figure}

Fig.~\ref{Fig:degreedistribution} illustrates the degree
distribution $p(k)$ of the visibility graph mapped from the energy
dissipation rate series in three-dimensional fully developed
turbulence with $10^{6}$ data sets. It is evident that the degree
distribution has a power-law tail
\begin{equation}\label{Eq:PDF:k}
 p(k)\sim k^{-\alpha}~,
\end{equation}
where the exponent is estimated to be $\alpha=3.0$. This
power-law distribution is obtained empirically using the method
proposed by Clauset, Shalizi and Newman
\cite{Clauset-Shalizi-Newman-2009-SIAMR}. The power-law exponent
is determined by maximum likelihood estimation based on the
Kolmogorov-Smirnov statistic
\cite{Clauset-Shalizi-Newman-2009-SIAMR} and the power-law
distribution is statistically significant. It means that the
visibility graph under investigation is scale-free. The power-law
degree distribution has been found in the visibility graphs mapped
from fractal series such as the Brownian motion and Conway series
\cite{Lacasa-Luque-Ballesteros-Luque-Nuno-2008-PNAS}.
Fig.~\ref{Fig:degreedistribution} shows that multifractal data sets
of the energy dissipation rate in turbulence also converts into
scale-free networks and the exponent of the degree distribution of
the network may depend on the correlation of the series.

The power-law degree distribution of the visibility
graph has implications for the fractality of the original time
series
\cite{Lacasa-Luque-Ballesteros-Luque-Nuno-2008-PNAS,Lacasa-Luque-Luque-Nuno-2009-EPL,Ni-Jiang-Zhou-2009-PLA}.
This is consistent with the well known result that the energy
dissipation rate series is statistically self-similar
\cite{Frisch-1996}. For nonstationary time series, such as stock
price indexes, fractional Brownian motions, Conway series, and
strike intervals, there is a linear dependence of the Hurst index on
the tail exponent
\cite{Lacasa-Luque-Luque-Nuno-2009-EPL,Ni-Jiang-Zhou-2009-PLA}.
However, no quantitative relation between the Hurst index of
stationary time series and the tail exponent of the degree
distribution has been established. Indeed, if we apply
rudely the relation in
Refs.~\cite{Lacasa-Luque-Luque-Nuno-2009-EPL,Ni-Jiang-Zhou-2009-PLA},
we will come to the conclusion that the energy dissipation rate
series is anticorrelated since the resulting Hurst index is less
than $1/2$, which is consistent with the empirical facts
\cite{Liu-Zhou-2009-XXX}.

\subsection{The visibility graph is not self-similar}

\begin{figure}[htb]
\centering
\includegraphics[width=7cm]{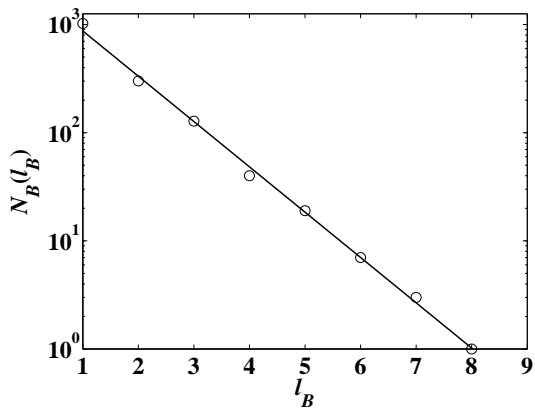}
\caption{\label{Fig:networkfractal} Edge covering box-counting
analysis of the visibility graph. $N_{B}$ and $l_{B}$ are in an
exponential relation, and the network is non-fractal.}
\end{figure}

The relation between scale-free networks and self-similar networks
has been discussed recently
\cite{Song-Havlin-Makse-2005-Nature,Song-Havlin-Makse-2006-NP,Goh-Salvi-Kahng-Kim-2006-PRL},
and fractal networks seem to exhibit scale-free features while
scale-free networks are not always self-similar. The fractal nature
of the network can be revealed by the well-known box-counting
method. Either the node-covering
\cite{Song-Havlin-Makse-2005-Nature} or the edge-covering
\cite{Zhou-Jiang-Sornette-2007-PA} box-counting method needs to
calculate the minimum number $N_{B}$ of boxes with size $l_{B}$ that
can cover all the nodes or edges of the network, where $l_{B}-1$ is
the maximum distance of all possible pairs of nodes in each box in
the node-covering box-counting method while $l_{B}$ represents the
largest distance in the edge-covering method. If the network is
self-similar, the minimum number $N_{B}$ of covering boxes scales
with respect to box size $l_{B}$ as
\begin{equation}
N_{B}\sim l_{B}^{-D}.
\end{equation}
where $D$ is the fractal dimension of the network. In order to
obtain the minimum number of the covering box, the simulated
annealing algorithm is implemented
\cite{Zhou-Jiang-Sornette-2007-PA} and the computation may be too
enormous for the network mapped from the whole series. In this work,
the graph mapped from the 1000 continuous points of the turbulence
series are considered to study the fractal feature.
Fig.~\ref{Fig:networkfractal} shows the results of the edge-covering
box-counting analysis of the visibility graph. As illustrated in
Fig.~\ref{Fig:networkfractal}, the number of boxes $N_{B}$ decreases
exponentially with the size of the boxes $l_{B}$, rather than a
power law. Thus, the visibility graph in our work is not fractal.

According to Song et al.
\cite{Song-Havlin-Makse-2005-Nature,Song-Havlin-Makse-2006-NP}, the
network with hubs (the high-degree nodes) attraction will not be
self-similar. Whether the interaction is attraction or repulsion
among the hubs of the network can be determined from the correlation
between the degrees of different nodes. The degree correlation can
be quantified by a correlation coefficient $r$
\cite{Newman-2002-PRL,Newman-2008-PT}:
\begin{equation}\label{Eq:correlation}
r  =\frac{\langle k_{1}k_{2}\rangle-\langle k_{1}\rangle \langle
k_{2}\rangle}{\sigma_{k}^{2}}
\end{equation}
where $\sigma_{k}^{2}$ is the variance of all the node degrees, and
$k_{1}$, $k_{2}$ are the degrees of the nodes at the ends of all the
edges.

It is clear that the degree correlations of adjacent nodes have
strong effects on the structure of the network. Networks with
positive correlation ($r>0$) tend to have a core-periphery
structure, and the hubs are attracted to each other with large
probability, while the hubs of a network with negative correlation
($r<0$) are repulsive. In our case, we obtain that $r=0.1204$ for
the network mapped from the whole time series. The positive
correlation of the whole network indicates that the graph mapped
from the energy dissipation rate series is hub-attractive and thus
not self-similar. According to the visibility graph method, it is
easy to find that large values of the energy dissipation rates
correspond to the hubs in the network with large probability, and
the points with large values can "see" each other easily which means
that edges exist between the hubs.

More precisely, the correlation coefficient $r$ in
Eq.~(\ref{Eq:correlation}) is not a good indicator to determine the
fractality of the network. As is explained in
Ref.~\cite{Song-Havlin-Makse-2006-NP}, for the property of a higher
degree of hub repulsion to be the hallmark of fractality, ``the
anticorrelation has to appear not only in the original network, but
also in the renormalized networks at all length scales.'' What is
important is not the degree correlation but the scaling of the
degree correlation \cite{Gallos-Song-Makse-2008-PRL}. In addition,
some measures of anticorrelation, such as the Pearson coefficient
$r$, cannot capture the difference between a fractal and a
non-fractal network, since $r$ is not invariant under
renormalization
\cite{Song-Havlin-Makse-2006-NP,Gallos-Song-Makse-2008-PRL}.
However, positive correlation of degrees using the Pearson
coefficient is sufficient to confirm that the network is not
fractal.

\subsection{Allometric scaling of the visibility graph}

The allometric scaling was first proposed in biology
\cite{West-Brown-Enquist-1997-Science,West-Brown-Enquist-1999-Science},
and it turns out that the various scalings are characterized by the
allometric equation:
\begin{equation}\label{ae}
Y\sim M^{b}
\end{equation}
where $Y$ is any physiological, morphological or ecological variable
that appears to be correlated with size or body mass $M$, and $b$ is
the scaling exponent. Allometric scaling
laws have been uncovered in many networks, such as organ metabolic
networks \cite{Santillan-2003-EPJB}, food webs
\cite{Garlaschelli-Caldarelli-Pietronero-2003-Nature}, world
trade/investment webs \cite{Duan-2007-EPJB,Song-Jiang-Zhou-2009-PA},
and so on \cite{Banavar-Maritan-Rinaldo-1999-Nature}. The allometric
scaling law is proposed in a network based on spanning trees. With
one root and plenty of branches and leaves, a tree can be considered
as directed from root to leaves. For each node $i$ in the tree, we
calculate $A_{i}$ of node $i$ that the summation of value $A$ of its
branches along with node $i$ itself
\begin{equation}
A_{i}=\sum_{j}A_{j}+1
\end{equation}
and the quantity $C_{i}$ of node $i$ is the sum of the $C$ values of its branches and $A_{i}$
\begin{equation}
C_{i}=\sum_{j}C_{j}+A_{i},
\end{equation}
where $j$ is the son node of $i$.

Imagining that there is a flow from the root to the leaves of the
tree, the value $A$ stands for the size of node while the value $C$
represents the transport rate through the node. According to
Eq.~(\ref{ae}), if the property of allometric scaling exists in the
tree, the rate $C_{i}$ and the size $A_{i}$ should depend on a
power-law relation:
\begin{equation}
C\sim A^{\eta}.
\end{equation}
The exponent $\eta$ is a very important parameter of the tree
structure. The transportation is more efficient for smaller $\eta$
values. The chain-like trees and star-like trees are the two
extremes of the spanning trees, and the value of the allometric
exponent $\eta$ is also between that of the two trees. It is easy to
obtain that $\eta=2^{-}$ for chain-like trees and $\eta=1^{+}$ for
star-like trees. It follows that $1<\eta<2$. We note that not all
trees exhibit allometric scaling properties
\cite{Jiang-Zhou-Xu-Yuan-2007-AIChE,Duan-2007-EPJB}, and different
spanning trees for the same network may exhibit different allometric
features.

For a network, there are various spanning trees with very different
structures. In this work, we consider the skeleton
\cite{Goh-Salvi-Kahng-Kim-2006-PRL,Kim-Noh-Jeong-2004-PRE} of the
network. The skeleton is a particular spanning tree which is formed
by the edges with largest betweenness centrality of the original
network \cite{Kim-Noh-Jeong-2004-PRE}. The skeleton of the
scale-free network is also scale-free with a different exponent
$\alpha$, and if the original network is self-similar, the skeleton
also exhibits self-similarity. To extract the skeleton of a network,
we first determine the shortest paths between all pairs of nodes,
and the number of times an edge appears in the shortest paths is
defined as the weight of the edge,
\begin{equation}
w(E)=\sum_{i,j}c(i,j,E),
\end{equation}
where $c(i,j,E)=1$ if the shortest path from node $i$ to $j$ through
edge $E$ and $c(i,j,E)=0$ otherwise. The node with the largest
degree is considered as the root of the spanning tree. As a first
step, we select an edge with the largest weight and add it to the
tree if its inclusion will not form any loops. Once all the nodes of
the original network are added into the tree, the skeleton is
obtained.

\begin{figure}[htb]
\centering
\includegraphics[width=7cm]{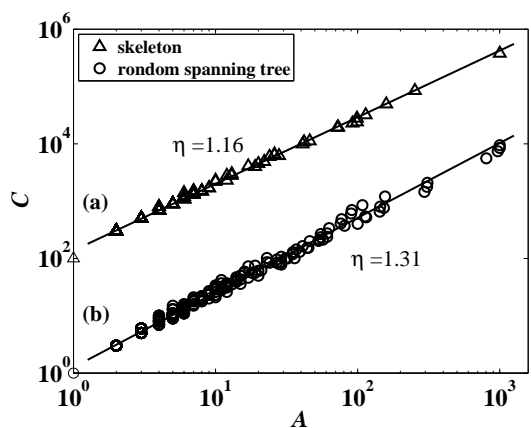}
\caption{\label{Fig:ascompares} Comparing the allometric scaling
behavior of the skeleton and a random spanning tree of the
visibility graph mapped from the same series. The lines are the best
power-law fits of the data. The data for the skeleton have been
shifted upward by two orders of magnitude for
clarity.}
\end{figure}

We randomly select 100 segments of the time series, each segment
having 1000 data points. The 100 sub-series are converted into
visibility graphs and their skeleton spanning trees are determined
accordingly. Fig.~\ref{Fig:ascompares} shows the comparison of the
scaling behavior between the skeleton and a random spanning tree.
The plot of skeleton is shifted upward for clarity and the lines are
the best fits of the data. The perfect power-law relations are
observed between $C$ and $A$ of all the considered segments, and the
line is the best fit of $\ln C$ against $\ln A$. The leaves of the
trees ($A=1$, $C=1$) are excluded from the fitting
\cite{Garlaschelli-Caldarelli-Pietronero-2003-Nature}. The exponents
of all the skeletons of different segments are around $1.16$ with
slight fluctuations, indicating that the allometric scaling property
of the visibility graph skeleton remains unchanged at different
parts of the turbulence time series. We find that $\eta=1.163 \pm
0.005$. We also observe that the data points of the random spanning
tree are more dispersive in the scatter plot than those of the
skeleton. It is interesting to note that this allometric scaling
exponent is quite universal in many different networks
\cite{Garlaschelli-Caldarelli-Pietronero-2003-Nature,Jiang-Zhou-Xu-Yuan-2007-AIChE}.
However, the physical implication and the relation to the dynamics
of the turbulence signal of this universal topological structure of
the visibility graph are unclear.

\section{Summary}
\label{S1:Conclusion}

The visibility graph algorithm is a new method for time series
analysis, which converts time series into complex networks. In this
way, the time series can be studied from a complex network
perspective. The visibility graphs inherit many properties of the
associated time series. We have investigated the visibility graphs
constructed from the time series of energy dissipation rate in
turbulence. We found that the visibility graph has a power-law tail
in its degree distribution and is thus scale free, which corresponds
to the self-similarity of the energy dissipation rate series. The
edge-covering box-counting method shows that the number of boxes
decays as an exponential function with respect to the box size,
which indicates that the visibility graph is not fractal. The
positive degree correlation coefficient $r$ and the visibility
algorithm for the turbulence data show hub-hub attraction (rather
than hub-hub repulsion) in the network, which explains why there is
no self-similarity. In addition, the allometric
scaling properties of the visibility graph are studied. We selected
skeleton as the particular spanning tree of the network, and the
results show that the skeleton exhibits excellent allometric
scaling. The networks mapped from various segments of
the series of the turbulence have the same scaling
exponent $\eta=1.163\pm 0.005$. However, the physical implication of
the allometric scaling is not clear. We hope that our analysis will
stimulate the study of turbulent signals from the perspective of
complex networks.

\bigskip
{\textbf{Acknowledgments:}}

This work was partially supported by the National Basic Research
Program of China (2004CB217703), the Program for Changjiang Scholars
and Innovative Research Team in University (IRT0620), and the
Program for New Century Excellent Talents in University
(NCET-07-0288).

\bibliography{E:/Papers/Auxiliary/Bibliography}

\end{document}